\documentclass{./aa}  

\usepackage{graphicx}
\usepackage{txfonts}
\usepackage{natbib}
\usepackage{tablefootnote}
\usepackage{hyperref}
\def\folio{\ifnum\pageno=1\nopagenumbers\else\number\pageno\fi}

\def\lax    {\ifmmode{_<\atop^{\sim}}\else{${_<\atop^{\sim}}$}\fi}
\def\gax    {\ifmmode{_>\atop^{\sim}}\else{${_>\atop^{\sim}}$}\fi}
\newbox\grsign      \setbox\grsign=\hbox{$>$} 
\newdimen\grdimen   \grdimen=\ht\grsign
\newbox\simgreatbox \setbox\simgreatbox=\hbox{\raise.5ex\hbox{$>$}\llap
                        {\lower.5ex\hbox{$\sim$}}}\ht1=\grdimen\dp1=0pt
\newbox\simlessbox  \setbox\simlessbox =\hbox{\raise.5ex\hbox{$<$}\llap
                        {\lower.5ex\hbox{$\sim$}}}\ht2=\grdimen\dp2=0pt

\def\simless {\mathrel{\copy\simlessbox }}

\def\dh {\phantom{$0$}}
\def\dd {\phantom{$00$}}

\newbox\grsign \setbox\grsign=\hbox{$>$} \newdimen\grdimen \grdimen=\ht\grsign
\newbox\laxbox \newbox\gaxbox
\setbox\gaxbox=\hbox{\raise.5ex\hbox{$>$}\llap
     {\lower.5ex\hbox{$\sim$}}}\ht1=\grdimen\dp1=0pt
\setbox\laxbox=\hbox{\raise.5ex\hbox{$<$}\llap
     {\lower.5ex\hbox{$\sim$}}}\ht2=\grdimen\dp2=0pt
\def\gax{\mathrel{\copy\gaxbox}}
\def\lax{\mathrel{\copy\laxbox}}
\def\boxit#1    {\vbox{\hrule\hbox{\vrule\kern3pt
                  \vbox{\kern3pt#1\kern3pt}\kern3pt\vrule}\hrule}}
\def\h      {\ifmmode{^{\rm h}}\else{$^{\rm h}$}\fi}
\def\m      {\ifmmode{^{\rm m}}\else{$^{\rm m}$}\fi}
\def\s      {\ifmmode{^{\rm s}}\else{$^{\rm s}$}\fi}
\def\decas    {\ifmmode{{\rlap.}{''}}\else{${\rlap.}{''}$}\fi}
\def\mum     {\ifmmode{\mu{\rm m}}\else{$\mu{\rm m}$}\fi}
\def\s      {\ifmmode{^{\rm s}}\else{$^{\rm s}$}\fi}
\def\deg      {\ifmmode{^{\circ}}\else{$^{\circ}$}\fi}
\def\as     {\ifmmode {\rlap.}$\,$''$\,$\! \else ${\rlap.}$\,$''$\,$\!$\fi}
\def\decsec  {\ifmmode {\rlap.}$\,$^{\rm s}$\,$\! \else ${\rlap.}$\,$^{\rm s}$\,$\!$\fi}\def\decs  {\ifmmode {\rlap.}$\,$^{\rm s}$\,$\! \else ${\rlap.}$\,$^{\rm s}$\,$\!$\fi}

\def\kms    {\ifmmode{{\rm km~s}^{-1}}\else{km~s$^{-1}$}\fi}

\def\Lsun   {$L_{\odot}$}

\def\rstar  {$r_{\star}$}

\def\Mspy   {\ifmmode {M_{\odot} {\rm yr}^{-1}} \else $M_{\odot}$~yr$^{-1}$\fi}
\def\Mdot   {\ifmmode {\dot M} \else $\dot M$\fi}
\def\mhd    {\ifmmode {n_{{\rm H}_2}} \else $n_{{\rm H}_2}$\fi}
\def\mhcd   {\ifmmode {N_{{\rm H}_2}} \else $N_{{\rm H}_2}$\fi}

\def\El      {\ifmmode{E_{\ell}}\else{$E_{\ell}$}\fi}
\def\beam    {\ifmmode{\theta_{\rm B}}\else{$\theta_{\rm B}$}\fi}
\def\mjyb   {\ifmmode {{\rm mJy~beam}^{-1}} \else{mJy~beam$^{-1}$}\fi}
\def\mujyb   {\ifmmode {\mu{\rm Jy~beam}^{-1}} \else{$\mu$Jy~beam$^{-1}$}\fi}
\def\Trot   {\ifmmode{T_{\rm rot}}\else$T_{\rm rot}$\fi}    
\def\Tvib   {\ifmmode{T_{\rm vib}}\else$T_{\rm vib}$\fi}    
    
\def\Teff   {\ifmmode{T_{\rm eff}}\else$T_{\rm eff}$\fi}

\def\ITRS   {\ifmmode{\smallint {\rm T}_{R}^{*}dv}\else{$\smallint 
{\rm T}_{R}^{*}dv$}\fi}
\def\ITRS   {\ifmmode{\smallint {\rm T}_{R}^{*}dv}\else{$\smallint 
{\rm T}_{R}^{*}dv$}\fi}
\def\ITAS   {\ifmmode{\smallint {\rm T}_{A}^{*}dv}\else{$\smallint 
{\rm T}_{A}^{*}dv$}\fi}

\def\hzo        {H$_2$O}

\def\lefttitle#1  {\noindent \hangindent=18.0pt \hangafter=1 {#1} \par}
\def\vol#1  {{\bf {#1}{\rm,}\ }}
\font\tenssb=cmssbx10
\font\tenbf=cmbx10
\font\sevenbf=cmbx8
\font\fivebf=cmbx6
\def\unetdemi    {\smallskipamount=6pt plus2pt minus2pt
                  \medskipamount=12pt plus4pt minus4pt
                  \bigskipamount=24pt plus8pt minus8pt
                  \normalbaselineskip=16pt plus0pt minus0pt
                  \normallineskip=2pt
                  \normallineskiplimit=0pt
                  \jot=6pt
                  {\def\smallskip {\vskip\smallskipamount}}
                  {\def\medskip   {\vskip\medskipamount}}
                  {\def\bigskip   {\vskip\bigskipamount}}
                  {\setbox\strutbox=\hbox{\vrule 
                    height17.0pt depth7.0pt width 0pt}}
                  \parskip 12.0pt
                  \normalbaselines}
\def\smallerspace {\smallskipamount=3pt plus0pt minus0pt
                  \medskipamount=6pt plus0pt minus0pt
                  \bigskipamount=10.5pt plus0pt minus0pt
                  \normalbaselineskip=10.5pt plus0pt minus0pt
                  \normallineskip=1pt
                  \normallineskiplimit=0pt
                  \jot=3pt
                  {\def\smallskip {\vskip\smallskipamount}}
                  {\def\medskip   {\vskip\medskipamount}}
                  {\def\bigskip   {\vskip\bigskipamount}}
                  {\setbox\strutbox=\hbox{\vrule 
                    height8.5pt depth3.5pt width 0pt}}
                  \parskip 0pt
                  \normalbaselines}
\def\memospace    {\smallskipamount=4pt plus1pt minus1pt
                  \medskipamount=6pt plus2pt minus2pt
                  \bigskipamount=14pt plus6pt minus6pt
                  \normalbaselineskip=14pt plus0pt minus0pt
                  \normallineskip=1pt
                  \normallineskiplimit=0pt
                  \jot=4pt
                  {\def\smallskip {\vskip\smallskipamount}}
                  {\def\medskip   {\vskip\medskipamount}}
                  {\def\bigskip   {\vskip\bigskipamount}}
                  {\setbox\strutbox=\hbox{\vrule 
                    height17.0pt depth7.0pt width 0pt}}
                  \parskip 2.0pt
                  \normalbaselines}
\def\memowidespace    {\smallskipamount=5pt plus1pt minus1pt
                  \medskipamount=7.5pt plus2pt minus2pt
                  \bigskipamount=17.5pt plus6pt minus6pt
                  \normalbaselineskip=17.0pt plus0pt minus0pt
                  \normallineskip=1.25pt
                  \normallineskiplimit=0pt
                  \jot=5pt
                  {\def\smallskip {\vskip\smallskipamount}}
                  {\def\medskip   {\vskip\medskipamount}}
                  {\def\bigskip   {\vskip\bigskipamount}}
                  {\setbox\strutbox=\hbox{\vrule 
                    height21.25pt depth8.75pt width 0pt}}
                  \parskip 2.5pt
                  \normalbaselines}
\bibpunct{(}{)}{,}{a}{}{,}
\def\raw {\rightarrow}

\def\dg {\phantom{$>9$}}

\def\dd {\phantom{$00$}}

\def\Mspy   {\ifmmode {M_{\odot} {\rm yr}^{-1}} \else $M_{\odot}$~yr$^{-1}$\fi}
\def\Mdot   {\ifmmode {\dot M} \else $\dot M$\fi}
\def\as     {\ifmmode {\rlap.}$\,$''$\,$\! \else ${\rlap.}$\,$''$\,$\!$\fi}
\def\hml      {\hbox{$(0,1^{{1}_{\rm c}},0)$}}
\def\hmld      {\hbox{$(0,1^{{1}_{\rm d}},0)$}}

\def\vs {\vspace{0.07cm}}
\usepackage{color}

\begin{document}

\title{Widespread HCN maser emission in carbon-rich evolved stars}%\thanks{Herschel is an ESA space
%observatory with science instruments provided by European-led
%Principal Investigator consortia and with important participation
%from NASA.}}

%\subtitle{}

 \author{K. M. Menten
 \inst{1}
 \and
 F. Wyrowski
 \inst{1}
 \and
D. Keller
\inst{1,2}\fnmsep
\thanks{Member of the International Max Planck Research School (IMPRS) for Astronomy and Astrophysics at the Universities of Bonn and Cologne.}\fnmsep     
\and
T. Kami\'nski
\inst{3}
%   \offprints{}
}
 \institute{   
%1
Max-Planck-Institut f{\"u}r Radioastronomie, Auf dem H{\"u}gel 69,
D-53121 Bonn, Germany 
\email{kmenten@mpifr.de}
\and
%2
Instituut voor Sterrenkunde, Katholieke Universiteit Leuven, Celestijnenlaan 200D, 3001
Leuven, Belgium
\and
Harvard-Smithsonian Center for Astrophyics,
60 Garden Street, Cambridge, MA 02138, USA
}

 \date{Received ... ; accepted}

\titlerunning{HCN maser emission in C-rich evolved stars}
% \abstract{}{}{}{}{} 
% 5 {} token are mandatory
\abstract
%Context
{HCN  is a major constituent of the circumstellar envelopes of carbon-rich evolved stars, and rotational lines from within its vibrationally excited states probe parts of these regions closest to the stellar surface. A number of such lines are known to show maser action. Historically, in one of them, the 177 GHz $J=2\rightarrow1$ line in the %\hml\textcolor[rgb]{0.988235,0.501961,0.0313726} 
$l$-doubled bending mode %\LEt{I'm not quite clear which way
%this ought to be written. Is this a double-bending mode, and
%the double bending occurs in (...)l? Then it should be "in the (...)
% double-bending mode". If it's a bending mode that is doubled
%in the (...)l  (line), it should be
%the (...)l-doubled bending mode". Please check this and change
%throughout as required}
has been found to show relatively strong maser action, with results only published for a single object, the archetypical high-mass loss asymptotic giant branch (AGB) star IRC+10216.} 
%Aim
{To examine how common 177 GHz HCN maser emission is, we conducted an exploratory survey for this line toward a select sample of carbon-rich asymptotic giant branch stars that are observable from the southern hemisphere.}
%Methods
{We used the Atacama Pathfinder Experiment 12 meter submillimeter Telescope (APEX) equipped with a new receiver to simultaneously observe three $J=2\rightarrow1$ HCN rotational transitions, the \hml\ and \hmld\ $l$-doublet components, and the line from the (0,\dh{0},\dh{0})  ground state.} %density and temperature of the emission region.
%Results
{The \hml\ maser line is detected toward 11 of 13 observed sources, which all show emission in the (0,\dh{0},\dh{0})  transition. In most of the sources, the peak intensity of the \hml\ line  rivals that of the (0,\dh{0},\dh{0})  line; in two sources, it is even stronger. Except for the object with the highest mass-loss rate, IRC+10216, the \hml\ line covers a smaller velocity range than the (0,\dh{0},\dh{0})  line. The \hmld\ line, which is detected in four of the sources, is much weaker than the other two lines and covers a velocity range that is smaller yet, again except for IRC+10216. Compared to its first detection in 1989, the profile of the \hml\ line observed toward IRC+10216 looks very different, and we also appear to see variability in the (0,\dh{0},\dh{0})  line profile (at a much lower degree).  Our limited information on temporal variability, taken together with results for another HCN maser line from the literature, disfavors a strong correlation of maser and stellar continuum flux.}
%Conclusions
{Maser emission in the 177 GHz $J=2\rightarrow1$ \hml\ transition of HCN appears to be common in carbon-rich AGB stars. Like for other vibrationally excited HCN lines, our observations indicate that the origin of these lines is in the acceleration zone of the stellar outflow in which dust is  forming. For all the stars toward which we detect the maser line, the number of photons available at 7 and 14 $\mu$m, corresponding to transitions to vibrationally excited states possibly involved in its pumping, is found to be far greater than that of the maser photons, which makes radiative pumping feasible. Other findings point to a collisional pumping scheme, however.}


 \maketitle
%
%________________________________________________________________

\section{\label{intro}Introduction: HCN in carbon-rich evolved stars and elsewhere}
%\section{\label{hcnspec}HCN in AGB stars: relevant spectroscopy and chemistry}
Hydrogen cyanide (HCN), one of the first molecules discovered by millimeter-wavelength astronomy \citep{Snyder1971}, is a common molecular species in  the interstellar medium (ISM), namely in star-forming regions. Moreover, HCN is also abundant  in the atmospheres and the circumstellar envelopes (CSEs) of carbon-rich asymptotic giant branch (AGB) stars. The substantial electric dipole moment of the HCN molecule, 2.99 D \citep{Ebenstein1984}, makes its pure rotational lines probes of the denser ISM and CSE environments.

In the atmospheres of cool C-rich AGB stars (``C-Miras''), stars whose abundance of carbon is higher than that of oxygen, after H$_2$ and CO, HCN becomes the third most abundant ``parent'' molecule \citep{Tsuji1964, Cherchneff2006}; \citet{Schoier2013} derived a median HCN/H$_2$ abundance ratio of $3\cdot10^{-5}$ for a sample of C-rich AGB stars.
Since the detection of its $J =1\rightarrow0$ ground-state line toward the archetypical high mass-loss very evolved C-rich AGB star IRC+10216 \citep[CW Leo, ][]{Morris1971}, rotational lines from levels with higher angular momentum quantum number, $J$, from the vibrational ground state have been detected, as well
as %\citep{Crosas1997}. Moreover, 
 a wide variety of rotational lines from excited vibrational states, %of HCN were found 
starting with the study of 
\citet{ZiurysTurner1986}. Most of these lines were found in the nearby IRC+10216 \citep[e.g., ][]{Avery1994, Groesbeck1994, Cerni2011}, but vibrationally excited HCN emission was also detected 
toward other C-rich AGB stars \citep[e.g., ][]{Bieging2001}, the more evolved protoplanetary nebula CRL 618 \citep{Thorwirth2003a}, in star-forming 
regions \citep[e.g., ][]{ZiurysTurner1986, Rolffs2011, Veach2013}, and even in the warm and dense core regions of starburst galaxies 
\citep{Salter2008, Sakamoto2010}.
%REFs Moreover, quite surprizingly, the molecule was also detected in oxygen-rich evolved objects. REFS

%In contrast to oxygen-rich mass-losing, cool evolved stars toward which maser emission is commonly observed in a plethora of lines from the SiO and \hzo\ molecules, only a few C-rich  
Strong maser amplification from molecules around C-rich evolved stars is much rarer than for O-rich stars, which very frequently host maser emission in a plethora of transitions from the SiO and \hzo\  molecules \citep[see, e.g., ][]{Gray2012, Gray2016}. Vibrationally excited HCN lines account for most of the known masing transitions found so far toward C-rich stars. The two most prominent of these consist of low %rotational quantum number, 
$J$ lines from within the two lowest vibrationally excited states of the \textit{l}-doubled %vibrationally excited 
bending mode with energies of $\approx 1030$  and 2060 K above the ground state \citep{Guilloteau1987,Lucas1988,LucasCernicharo1989}. %The %state with rotation upper state quantum  numbers 1, 2 and 3  

\citet{Bieging2001} gives a summary of stellar HCN masers and also reports the first detection of maser emission in the $J=3\rightarrow2$ and $4\rightarrow3$ transitions of the \hml\  state. All the stars hosting such masers appear to have relatively high mass-loss rates on the order of $10^{-6}$ to $10^{-5}$ \Mspy. From much more highly excited states, laser emission was detected in the $J=9-8$ rotational line from the third overtone of a bending 
mode state, $(0,4^0,0)$, and also in the $J = 10-9$ lines between the Coriolis-coupled $(1,1^1,1)$ and 
$(0,4^0,0)$ states \citep{Schilke2000, SchilkeMenten2003}. These lines have upper-level energies 
around 4200 K.  At the other extreme, evidence for maser emission has been 
reported in the lowest frequency ($J=1\rightarrow0$) rotational line of the vibrational ground state toward optically bright 
``blue'' carbon stars  \citep{Izumiura1995, Olofsson1998} that have much lower mass-loss rates than the values quoted above.

A relatively strong ($\sim 400$~Jy) HCN maser line connecting
the  $J =2$ and 1 rotational levels within the 
\hml\ ($ \equiv$ \textit{v}$_2 = 1_{\rm c}$)  \textit{l}-doubled vibrationally excited bending mode state of HCN\footnote{See Sect. \ref{hcnspec} for information on HCN spectroscopy.} was discovered toward IRC+10216 by \citet{LucasCernicharo1989}. Given its elevated lower level energy (1029~K), it is 
clear that this line %, from the \hml\ state, 
arises from the inner regions of the CSE of this object.  In a \textit{\textup{Note added in proof}} to the Letter reporting their discovery, \citet{LucasCernicharo1989} mention that ``a preliminary search in similar objects has led to six detections''. 

In 2015, the new SEPIA receiver was  commissioned on the Atacama Pathfinder Experiment 12 meter telescope, APEX\footnote{This publication is based on data acquired with the Atacama Pathfinder Experiment (APEX). APEX is a collaboration between the Max-Planck-Institut fur Radioastronomie, the European Southern Observatory, and the Onsala Space Observatory.}. It covers the 157--212 GHz range and thus, near 177 GHz, the frequencies of three HCN $J=2\rightarrow1$ rotational lines, namely the \hml\ maser transition, its %non-masing \hmld\  
counterpart from the other \textit{l}-doublet component, and that from the vibrational ground state. All three can be observed simultaneously.

Here we report on an exploratory survey for these three lines toward a sample of (mostly) southern C-rich AGB stars. In  Sect. \ref{sample} we introduce our observed sample. In Sect.  \ref{obs} we describe the observations made with the APEX telescope and present the data we obtained. This is followed by a brief  general description of HCN spectroscopy relevant to our study (Sect. \ref{hcnspec}). Our results are presented in Sect. \ref{secresults} and discussed in Sect. \ref{discussion}, where special emphasis is given to IRC+10216. We summarize our results and present an outlook on future studies in Sect.\ref{summary}.
% we give a g and the molecule's occurrance in CSEs.  Thereafter, in Sect.  \ref{obs}, Out spectra are discussed in the context of our program sources in \ref{discussion}. We draw a summary and give an  outlook in \ref{summary}

% the a Given the SEPIA receiver's 4 GHz IF bandwidth a total of 7 other $J =2 -1$ with level energies between 4 and 
%4700 K can be observed simultaneously with the maser line (see Table). We note that HCN rotational lines from similarly 
%high energy levels have been previously observed toward IRC+10216 with the CSO 10.4 meter telescope (see Fig. 3 of 
%Schilke et al. 2000, ApJ 528, L37). 

\section{\label{sample}Sample of C-rich carbon stars}
Our sample is presented in Table \ref{tab:sample}. For this first APEX survey, we made a ``success-oriented'' selection of 
C-rich AGB stars that are known to have HCN emission in the vibrational  ground-state $J=1\rightarrow0$ line as listed in the catalog of \citet{Loup1993}. For most sources we present two estimates for their distance: the first,  $D^{\rm BF}$, comes from \citet{Loup1993}, who compiled or estimated bolometric fluxes for a sample of AGB stars and assumed, uniformly, a luminosity of $10^4$~\Lsun. The second,  $D^{\rm PLR}$, was obtained from the catalog published by \citet{Menzies2006}. These values were obtained using the well-defined C-Mira period-luminosity relation established by \citet{Feast1989} and \citet{Whitelock2003} and are mostly based on a new dataset presented by \citet{Whitelock2006}. For some stars, period and phase were derived from visual data available through the web site of AAVSO's International Variable Star Index\footnote{\url{https://www.aavso.org/vsx/index.php}}.  Details on the phase determinations are given in the appendix. When both distances are available, they  agree quite well for
most of the objects, that is, they agree to within 30\%. %\ of the 
Because of the rather simplistic assumption of a uniform luminosity made by Loup et al., in this paper $D^{\rm PLR}$ is used for the distance when available. If it is not, we use $D^{\rm BF}$. 
Published values of the stellar LSR velocities, $\varv$$_{\rm LSR}^{*}$, and the terminal velocities of their CSEs, $\varv$$_{\inf}^{*}$, were again taken from the \citet{Loup1993} catalog or from \citet{Olofsson1993a}. In both papers these values were determined from spectra of the CO $J = 1\rightarrow0$ and $2\rightarrow1$ lines (and also from HCN $J = 1\rightarrow0$ spectra in the former). For the stellar mass-loss rates, \Mdot, we present a range of values that envelop the numbers given by \citet{Loup1993} and \citet{Olofsson1993a}. These values are based on modeling of CO line data. A discussion of the large spread in \Mdot\ is beyond the scope this paper, in which mass-loss rate is only invoked in a qualitative discussion of the propensity to find maser emission as a function of this quantity.% In some case %Except were noted, the two values for the mass-loss rates are also taken from \citet{Loup1993}. The first value is based was calculated by these authors themselves, while the second one is the result of more sophisticated modeling by \citet{Kastner1990}. 

\begin{table*}
\caption{\label{tab:sample}Stellar sample information}
%\begin{center}
%\setlength{\tabcolsep}{0.06cm} 
\centering
%`\begin{tabular}{lcclcccccccc}
%\begin{tabular}{llccllrr}
%\begin{tabular}{S[table-format=4.1]}{llccllrrcrr}
%\begin{threeparttable}
%\begin{tabular}{lllccllrccrcc}
\begin{tabular}{llccllrccll}
%\begin{tabular}{llcclld{7.3}ccll}
\hline \hline

Object          & IRAS       &\multicolumn{2}{c}{Position}
                                           
                                                                      &$D^{\rm BF}$&$D^{\rm PLR}$ &$\varv$$_{\rm LSR}^{*}$&$2\varv$$_{\inf}^{*}$&\Mdot%\multicolumn{2}{c}{\Mdot}
                                                                      %&$S_{12\mu\rm m}$
                                                                      &\multicolumn{1}{c}{Period}&$\phi_{\rm IR}$\\%&$\phi$\\%$(\nu_1,\nu_2,\nu_3)$& $\varv$$_{\rm LSR}$-range &\multicolumn{1}{c}{$\int T_{\rm MB}d\varv$} \\%\Mdot$&$\varv$$_\infty$
                                                                                 %        & $\varv$$_{\rm LSR}$ range 
                                                                                    %        &$\int T_{\rm MB}d$\\%$\varv$&$\Mdot$$^{\rm Mod}$&$r_{\rm out}$&$\theta_{\rm out}$&$X_{{\rm NH}_3}$\\

                           & &$\alpha_{\rm J2000}$ 
                                              & $\delta_{\rm J2000}$         &\multicolumn{2}{c}{(kpc)}&\multicolumn{2}{c}{(\kms)}&%\multicolumn{2}{c}{(
                                              $(10^{-7}$\Mspy)&(d)\\%&(km~s$^{-1})$    &\multicolumn{1}{c}{(K km~s$^{-1})$} \\
                                                                                  
                                                                                   %&(\Mspy)                            
                                                                               %           &(km~s$^{-1}$   &(km~s$^{-1})$    &(K km~s$^{-1})$\\ &(\Mspy)&(cm) &($''$)\\
%\noalign{\smallskip\hline
%\noalign{\smallskip}
\hline
R For                                    & 02270$-$2619 & $02^{\rm h}29^{\rm m}15\decsec3$ & $-26^\circ05' 56''$&0.67&0.70&  $-$2&     38&9--22 &385 & 0.0\\        
R Lep                                   & 04573$-$1452 & 04 59 36.4  &$-$14 48 22     &0.48&0.47&   $+$16&  38           &5--20                                 &438 & 0.5\\
AI Vol                                   & 07454$-$7112 & 07 45 02.4  &$-$71 19 46      &0.75&0.83&   $-$39  &  27          &51                                     & 511 & 0.1\\
X Cnc                                   & 08525$+$1725  &08 55 22.9  &$+$17 13 52     &0.69&--&       $-$15  &  20         &0.4--8                               &180$^{\rm a}$& 0.5$^{\rm a}$\\
%X Cnc                                   & 08525$+$1725  &08 55 22.9  &$+$17 13 52     &0.69&--&       $-$15  &  20         &0.4--8                               &180$\tablefootnote{xxxtest}$& 0.5$^{\rm c}$\\
CQ Pyx (RAFGL 5254) & 09116$-$2439     &09 13 53.9  &$-$24 51 25    &0.42&1.14&  $0$  &26               &23--120                             &659 & 0.6\\
CW Leo (IRC+10216) & 09452$+$1330    &09 47 57.4  &$+$13 16 44    &0.12&0.14$^{\rm b}$& $-$26.5 &36          &200--400                           &630$^{\rm c}$&0.23\\
X Vel                                 &09533$-$4120      &09 55 26.9  &$-$41 35 15   &0.67&--&      $-$17&   17            &0.5--3.5                              &--&--\\    
RW LMi (CIT 6)           &  10131$+$3049   &10 16 02.3  &$+$30 34 19 &0.38&0.46& $-$2    &     30          &26--140                             &617&0.7\\       
U Hya                               &10350$-$1307                 &10 37 33.3  &$-$13 23 05   &0.35&--& $-$31& 18                  &1.1--5                      &183$^{\rm a}$ & 0.8$^{\rm a}$\\
V Hya                              &10491$-$2059               &10 51 37.3  &$-$21 15 00   &0.33&--&     $-$16 &22        &31--35                               &531&0.3\\
V358 Lup (RAFGL 4211) &15082$-$4808     &15 11 41.9  &$-$48 20 01    &0.67&0.95&$-$3    &    42            &93                                     &632&0.8\\        
X TrA                               &15094$-$6953                            &15 14 19.0 &$-$70 04 45    &0.47&--& $-$3&18   &0.4--1.6                            &320&--\\
II Lup$^{\rm d}$               &15194$-$5115 &15 23 05.7 &$-$51 25 59 &0.47&0.64&$-$15&46    &100$^{\rm e}$                                                  &576 & 0.4\\
%\noalign{\smallskip
\hline
%\noalign{\smallskip}
\end{tabular}
%\end{center}
%\footnotes{
\tablefoot{
Information on our sample of observed stars. Columns are (from left to right) the star name (alternative name), IRAS point source catalog designation, J2000 right 
ascension and declination, distance derived from bolometric flux, distance derived from Mira period-luminosity relation, stellar centroid LSR velocity, twice the terminal velocity (i.e., the full velocity range covered by spectral lines) of the star's CSE, %the flux density in the 12 $\mu$m band of the IRAS satellite (with power of 10 in parentheses; taken from the IRAS Point Source Catalog\tablefootnote{https://heasarc.gsfc.nasa.gov/W3Browse/all/iraspsc.html}), 
range of mass-loss rates (see Sect.\ref{sample}), variability period, and the infrared phase at the time of our APEX observations, i.e., 2015 May 28. For AGB stars the IR light has its maximum ($\phi_{\rm IR} = 0$) at a visual phase of $\approx0.1$. The periods and phases are based mainly on data in \citet[][see Sect.\ref{sample} and also our appendix]{Whitelock2006}. %In further columns we present the results of our APEX observations of the three $J=2-1$ lines in three rows for each star: preceded by the vibrational quantum numbers, we list the LSR velocity range covered by emission and the velocity-integrated main-beam brightness temperature for which the uncertainty in the last digit is given in parentheses. Upper limits in this quantity for the \hmld\ line were calculated by assuming that it covered the same velocity range as the \hml\ line if detected, otherwise that of the (0,\dh{0},\dh{0}) line was assumed, as was for upper limits in the \hml\ line. 
All our positions agree to within $2''$ with the 2MASS positions
of the stars, which themselves have an absolute accuracy of better than $0\as1$ \citep{Cutri2003}. For our discussion, the $D^{\rm PLR}$ value  is adopted for the distance, when available. In other cases we use $D^{\rm BF}$. The LSR and terminal velocities were taken from \citet{Loup1993}, who in general compile multiple literature values for these quantities mostly based on CO $J=1\rightarrow0$, $2\rightarrow1$ or HCN $J=1\rightarrow0$ spectra. Our adopted values are averages of the higher-quality entries and should have uncertainties $< 2$~\kms. % Periods are taken from \citet{Olofsson1993a} and are heterogeneous values (visual, photographic, blue; see their Sect. 2.2). %Except were noted, the two values for the mass-loss rates are also taken from \citet{Loup1993}. The first value is based was calculated by these authors themselves, while the second one is the result of more sophisticated modeling by \citet{Kastner1990}. 
%\hline
$^{\rm a}$Period and phase derived from visual data available through the web site of AAVSO's International Variable Star Index. %\tablefootnote{\url{https://www.aavso.org/vsx/index.php}}.  
$^{\rm b}$\citet{Menten2012} proposed a distance of 0.13 kpc for this star. $^{\rm c}$Combined value from $JHKL$ IR light curves \citep{Menten2012}.$^{\rm d}$For this star, its IRAS name is frequently used in the literature.  $^{\rm e}$\citet{Ryde1999}. 
}
%\end{threeparttable}
\end{table*}

\section{\label{obs}Observations} % and HCN spectroscopy summary}
Our observations  were made on 2015 May 28 under excellent weather conditions with the  12 m diameter APEX submillimeter telescope \citep{Gusten_etal2006} under project number M-095.F-9544A-2015. The amount of precipitable water vapor in the atmosphere as monitored by the APEX radiometer\footnote{\url{http://www.apex-telescope.org/weather/}} was 0.9 mm in the atmosphere above the Llano de Chajnantor.

We used the new prototype receiver developed for frequency band 5 of the Atacama Large Millimeter/submillimeter Array (ALMA). At the APEX telescope, it is a component of SEPIA, the Swedish-ESO PI receiver for APEX \citep{Billade2012, Immer2016}. This  is a dual-polarization sideband-separated (2SB) 
receiver that covers the frequency range 157.36--211.64 GHz. In particular, the central part of this band has been relatively little explored in the past because of strong atmospheric absorption 
caused by the $3_{13}-2_{20}$ line of para-\hzo\ near 183 GHz. At the excellent
site at 5100 m at which ALMA and APEX are located, the transmission in this band is quite acceptable for most of the time, and 
maser emission in this very \hzo\ line has previously been observed with APEX toward a variety of astronomical sources \citep{Immer2016, Humphreys2017}.

Each sideband of the receiver has a total bandwidth 
of 4 GHz. The central frequencies of the two sidebands are separated by 12 GHz, 
corresponding to an intermediate frequency (IF) band of 4--8 GHz. We centered the signal IF band at 177.780 GHz, a frequency close to  our main target lines  listed in Table\,\ref{lines}. 
For our observations we wobbled the telescope subreflector with a rate of 1.5 Hz between positions that were symmetrically offset in azimuth by $60''$, larger than any plausible HCN emission distribution (see Sect. \ref{cwprofile}).
Total observing times ranged from (mostly) $\sim 0.5$ h up to 2 h for the weaker sources. %(ON+OFF source) per source, except for R For an X Vel, which were observed twice this long.
%KMM: THIS IS FROM THE HEADERS. IS IT OFF OR ON+OFF TIME? 
Calibration was obtained using the chopper-wheel technique under consideration of the different atmospheric opacities in the signal and image sidebands of the 
employed 2SB receiver. % (see Fig. \ref{transmission}). 
The radiation was analyzed with the newest implementation of the MPIfR-built fast Fourier transform spectrometer \citep{Klein_etal2006}, which accounted for 
redundant overlap between sub modules and provided 52430 frequency channels  over the 4\,GHz wide intermediate frequency bandwidth with a channel spacing of 76.3 kHz, corresponding to 0.13 \kms.
To increase the signal-to-noise ratio, S/N, the spectra were smoothed to effective velocity resolutions appropriate for the measured widths even of narrow features, that is, 
$\sim0.5$--1 \kms. 

To check the telescope pointing, we used either the signal from the HCN $J=2\rightarrow1$ lines from the program stars themselves or that from the SiO $\varv = 1, J=4\rightarrow3$ maser line (172481.1175 MHz) from an O-rich AGB star close in the sky to one of our C-rich program stars. % to which showed strong emission in all of our sources, and 
Five-point crosses centered on the stellar position with half-beam
width offsets in elevation and azimuth were measured. Pointing corrections were derived from these measurements. The pointing
was found to be accurate to within $\approx3''$, acceptable
given the full-width at half-maximum (FWHM) beam size, $\theta_{\rm B}$, which is $36''$ at 177 GHz. We established a  main-beam brightness temperature, $T_{\rm MB}$, %$\int T_{\rm MB}d\varv$, 
scale (in K) by extrapolating the main-beam efficiencies,   $\eta_{\rm MB}$, from higher-frequency values observationally, determined by \citet{Gusten_etal2006}, to 177 GHz. We assumed $\eta_{\rm MB} = 0.7$. This value was validated by observations of Uranus. Our $T_{\rm MB}$ values can be multiplied by a factor of 33.4 to convert them into flux density units (in Jy).

\section{\label{hcnspec}HCN vibration-rotational spectroscopy}
\subsection{HCN vibrational modes}
The HCN molecule has three fundamental vibrational states: the CH stretching mode, designated $\nu_1$ [or (1,0,0)], the doubly degenerate bending mode, $\nu_2$ [or (0,1,0)], and the CN stretching mode, $\nu_3$ [or (0,0,1)]. The wavelengths of the fundamental
transitions of these modes are 3.0, 14.0, and 4.9 $\mu$m, and
they correspond  to temperatures of 2931, 1025, and 4841 K \citep{AdelBarker1934}. 
Because of the Earth's atmosphere,  direct ro-vibrational lines cannot easily be observed from the ground. %To probe the inner parts of IRC+10216's CSE,
\citet{Cerni1999} used the Short Wavelength Spectrometer (SWS) on board the Infrared Space Observatory (ISO) to study such lines from the stretching and the bending modes, including overtone and combinations bands toward IRC+10216. %Toward the same star, pure rotational lines from a variety of vibrational states, including overtones of the bending mode (0,2,0--0,8,0) and combination bands (e.g., 1,1,1 or 0,1,1)  were observed with the ISO Long Wavelength Spectrometer (LWS)  with upper levels J's, $J_{\rm u}$, of 18--48 \citep{Cerni1996} and with the IRAM 30 meter telescope ($J=3-2$, \citep{Cerni2011}).
%, even strong laser emission in the 

Each vibrational state contains a ladder of rotational levels, characterized by the angular momentum quantum number $J$.  Rotational  ($J \rightarrow J-1$) transitions can be observed from the ground for 
many $J$ between  $1$ and $12$. Compared to the IR bands mentioned above, these lines can be studied %from the ground 
with very high spectral resolution ($> 10^5$), \textit{\textup{and}} their emission can also be imaged with sub-arcsecond angular resolution and superb sensitivity with interferometers such as the IRAM NOrthern Extended Millimeter Array (NOEMA) and ALMA.

\subsection{\label{ltype}\textit{l}-type doubling of the $\nu_2$
bending-mode levels}
Since H--C--N can bend in two orthogonal directions, the $\nu_2$ vibrationally excited bending mode is doubly degenerate.
When the molecule is rotating and bending simultaneously, that
is, $J > 0$, this degeneracy is lifted, resulting in \textit{l}-type doubling with every 
rotational level being split into two sub-levels \citep{Nielsen1950}. The separation between these sub-levels increases with $J$ and is given by 
\begin{equation}
\label{eq:eq1}
\Delta \nu_J = \Delta \nu_{J=1}  \times J(J+1)/2,
\end{equation}
where in the $\nu_2 =1$ state, $\Delta  \nu_{J=1}$ = 448.9430 %and 1346.7652 
MHz (in frequency units) for the $J = 1$ %and 2 rotational 
level%, respectively,
; see \citet{MakiLide1967} and updated values listed in the Cologne Database for Molecular 
Spectroscopy, CDMS \citep{Mueller2005}\footnote{\url{http://www.astro.uni-koeln.de/cdms/catalog}}.
%It increases with $J$ as $\Delta \nu 
%Thus the two \textit{rotational} lines are separated in frequency by the amount given by  \ref{eq:eq1}.  
The  line connecting the lower $(0,1^{1_{\rm c}},0)$ components of the $\nu_2 =1$ doublet has an $\approx J\times11$ MHz lower frequency than the (0,\dh{0},\dh{0}) 
line with the same $J$, while the $(0,1^{1_{\rm d}},0)$ line, which connects the upper levels, has a frequency that is higher than that of the latter line by 
the amount given by the difference of Eq.  \ref{eq:eq1} evaluated, in our case, for $J=2$ and $J=1$. 
Note that several studies use a different nomenclature for the vibrationally excited \textit{l}-doublet states: $1^{1_{\rm e}}$ and $1^{1_{\rm f}}$ instead of $1^{1_{\rm c}}$ and $1^{1_{\rm d}}$ for the lower and upper sub-levels, respectively. For consistency with previous work on the \hml\ line, we here adhere to the latter convention.

%The  \hml\ %$(0,1^{1_{\rm c},0)$ 
%line connects the upper levels of the $l$-doubling split $J=2$ and 1 rotational states and the \hmld\ %$(0,1^{1_{\rm d},0}$ 
%line their lower levels.

%u_{J = 1} }
This means that the two lowest rotational transitions in the  $\nu_2 =1$ vibrationally excited state, the $J = 2\raw1$
lines, are separated by 897.8222 MHz and can thus be observed simultaneously and together with the (0,\dh{0},\dh{0})  line 
within the broad bandwidth provided by the APEX FFTS; see Sect. \ref{obs}. The rest frequencies of these three lines and their lower rotational level energies were taken from the CDMS and are given in Table \ref{lines}.

%Cologne Database for MolecularSpectroscopy \citep[CDMS,][]{Mueller2005}\tablefootnote{\url{http://www.astro.uni-koeln.de/cdms/catalog}}.

All three lines discussed here display  hyperfine structure (hfs); each is split into six components. Frequencies and other information on the hfs components can also be obtained from the CDMS (from links on the explanatory web pages for the HCN entries).  The relative intensities of the components and their offsets in velocity (relative to the values calculated from the ``centroid'' frequencies given in Table \ref{lines}) are displayed for the case of IRC+10216 in Fig. \ref{comp}. They are relevant for our discussion of the line widths of individual maser features in Sect. \ref{linewidths}.

\begin{table}[t]
%\begin{center}
\caption{\label{lines}HCN $J=2\rightarrow1$ rotational lines observed with SEPIA}
\centering
%\begin{threeparttable}
\begin{tabular}{clr}
 \hline \hline
%\multicolumn{7}{c}{Water lines (\nuteo)}\\
Transition &Frequency     &$E_\ell/k$\dh\\
%$J'_{K'_{a}K'_{c}} - J''_{K''_{a}K''_{c}}$
$(\nu_1,\nu_2,\nu_3)$                                   & \dh\dh(MHz)              &   (K)\dh \\
\hline
$(0,1^{1_{\rm c}},0)$ & 177238.6556(4)   &  1028.7 \\%714.9356\\
(0,\dh{0},\dh{0})                 & 177261.1112(3)    &   4.3\\ 
 $(0,1^{1_{\rm d}},0)$ & 178136.4778(4)   &  1028.3\\
\noalign{\smallskip}
 \hline
 \noalign{\smallskip}
 \end{tabular}
%\end{center}
\tablefoot{
Columns are (from left to right) vibrational %and rotational 
quantum numbers, %of upper and lower state, 
frequency (with last-digit uncertainty in parentheses) and energy above ground of the lower state in Kelvin; $k$ is the Boltzmann constant.
Frequency values were  taken from the CDMS. %Cologne database for Molecular Spectroscopy \citep{Mueller2005}\tablefootnote{\url{http://www.astro.uni-koeln.de/cdms/catalog}}.
The highly accurate values for the $\nu_2=1$ state lines are based on the molecular constants determined by \citet{Thorwirth2003b}.
%have formal uncertainties of order 50 kHz. More accurate values from a fit to the \hzo\ spectrum have been presented by \citet{Chen2000}. The difference between their values and  the ones used by us is typically of order 20 kHz, corresponding to less than 0.02 \kms, which is smaller than the uncertainties in our velocity determinations.
}
%\end{threeparttable}
\end{table}

\begin{figure*}[h]
%\centerline{\resizebox{1.0\hsize}{!}{\includegraphics[angle=0]{fournew.eps}}}
\centerline{\resizebox{0.75\hsize}{!}{\includegraphics[angle=0]{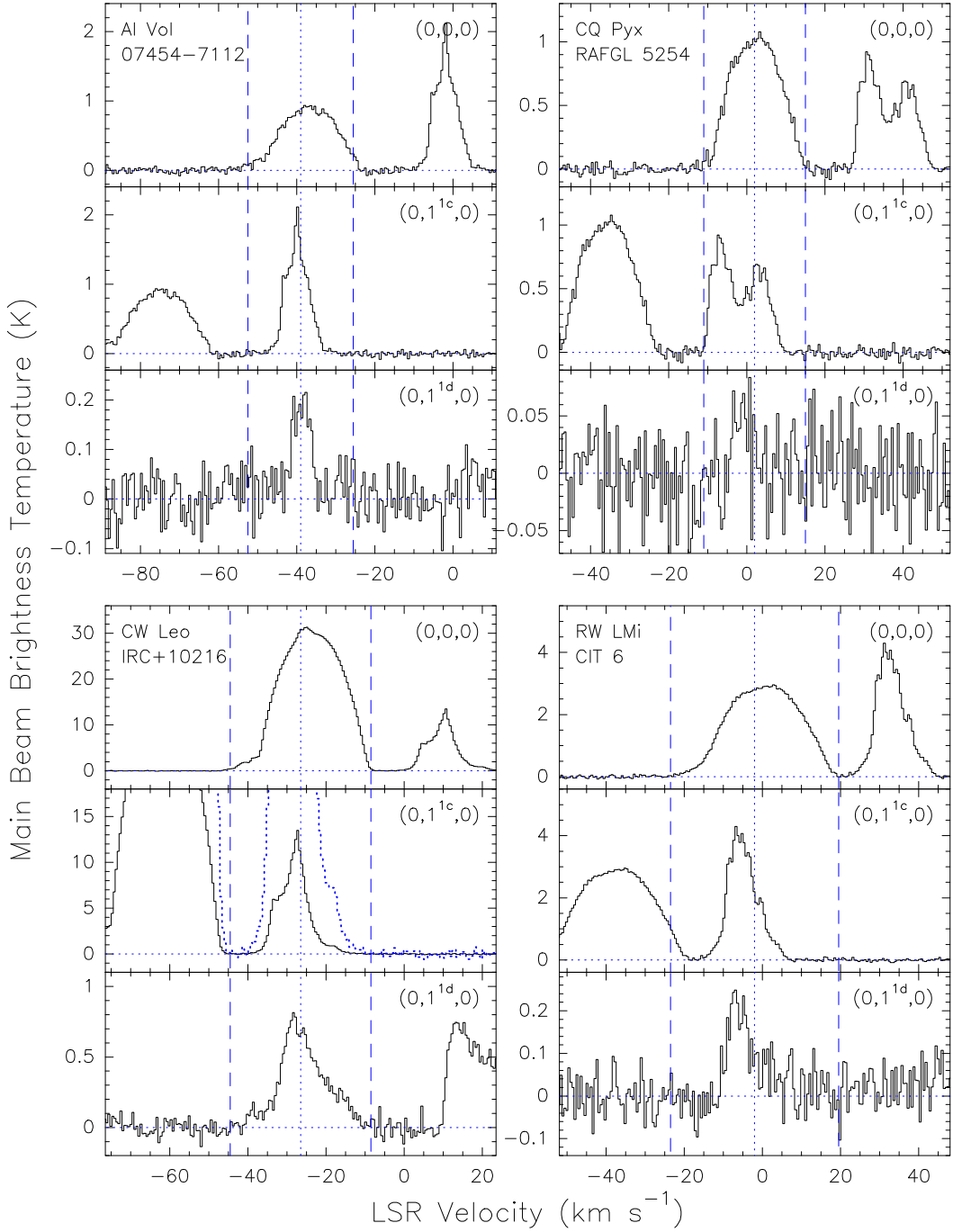}}}
\caption{APEX spectra of the four sources toward which all three HCN $J=2\rightarrow1$ lines are detected. For each source, the upper, middle, and lower panels show the spectra for the (0,\dh{0},\dh{0}) , \hml, and \hmld\ lines, respectively. 1 K $T_{\rm MB}$ corresponds to a flux density of 33.4 Jy. To facilitate the comparison between the lines, we used the same intensity scale for the first two lines, except for CW Leo. In addition, for CW Leo, the $T_{\rm MB}$ scale of the dotted blue line shows the base of the \hml\ spectrum scaled up by a factor of 8 compared to the ordinate in that panel. To adequately display the weak \hmld\ emission, a compressed $T_{\rm MB}$ scale is used for all sources for this line. The vertical dotted blue line marks the stellar velocity, while the dashed blue lines indicate the terminal velocity. The spectral line partially appearing in the \hmld\ spectrum of IRC+10216 at velocities $> 10$~kms\ is part of a multiplet component of the $N = 18\rightarrow17$ transition of C$_3$N.}
\label{four}
\end{figure*}

\section{\label{secresults}Results}
\begin{figure*}
\centerline{\resizebox{0.55\hsize}{!}{\includegraphics[angle=0]{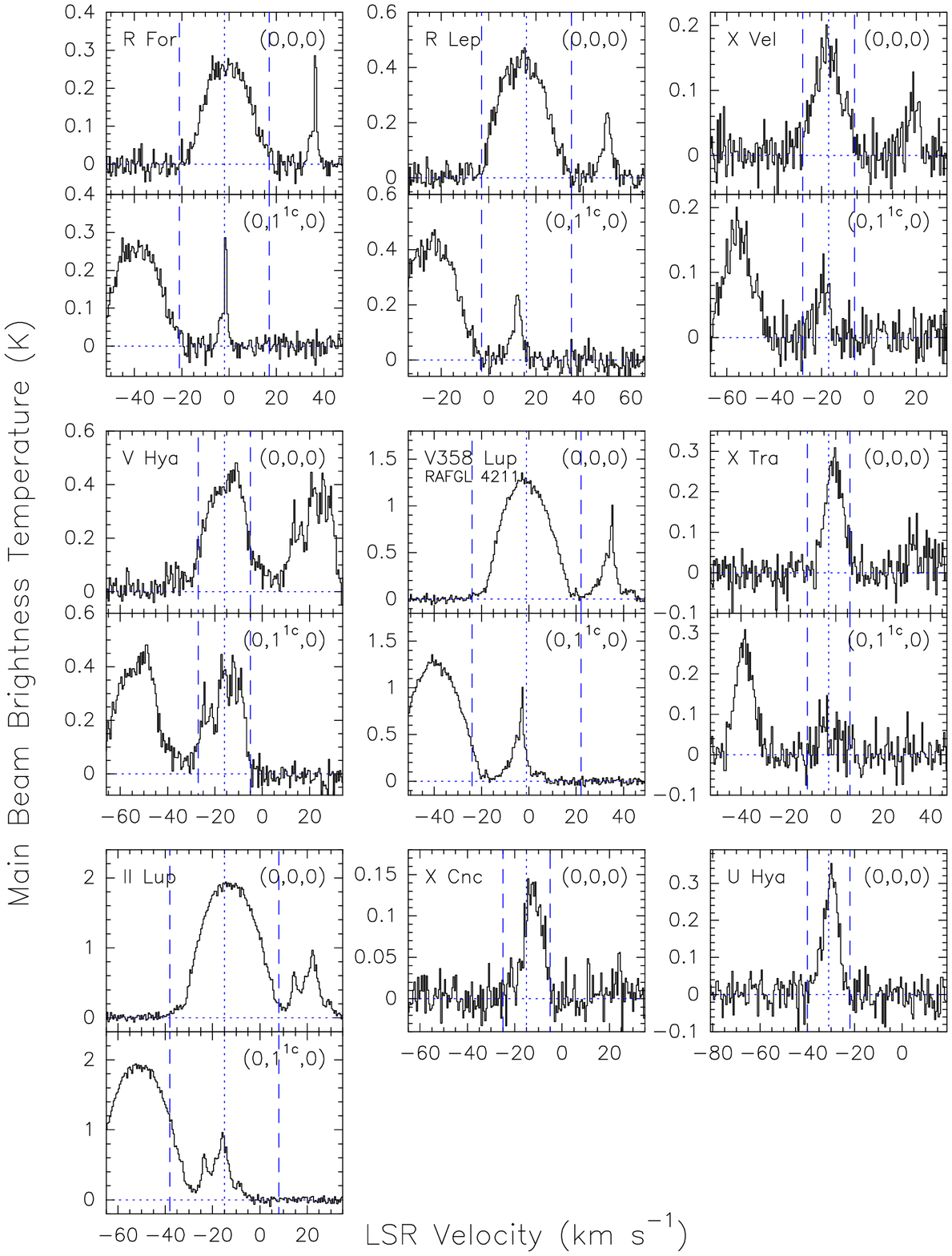}}}
\caption{APEX spectra of the seven sources toward which the (0,\dh{0},\dh{0})  and the \hml\  HCN lines are detected.  For each source, the upper and lower panels show the spectra for the (0,\dh{0},\dh{0})  and the \hml\ lines, respectively. Toward X Cnc and U Hya, only the (0,\dh{0},\dh{0})  line could be detected, and its spectra are shown at the center and right in the bottom row. The vertical dotted blue line marks the stellar velocity, while the dashed blue lines indicate the terminal velocity. A $T_{\rm MB}$ of 1 K corresponds to a flux density of 33.4 Jy.}

\label{rest}
\end{figure*}

Figures \ref{four} and \ref{rest} present all our observed spectra that show significant HCN emission. In Table \ref{results} we list for all three observed HCN $J=2\rightarrow1$ lines the full-width at  zero power (FWZP) velocity range over which significant emission is detected, our measured velocity-integrated main-beam brightness temperatures (the line fluxes), or upper limits for this quantity. We also list the ratio, $R$, of the (0,\dh{0},\dh{0})  line flux to that of the vibrationally excited lines (or its lower limit).
%\end{document}

The HCN (0,\dh{0},\dh{0})  line is detected in all of the 13 objects that we observed.
We find generally good agreement between the total velocity ranges,
that
is, twice the terminal velocity ($2\varv$$_{\inf}^{*}$), that we measure for this line and values found in the literature, which are 
mostly based on CO $J=1\rightarrow0$ and $2\rightarrow1$ line data (see Sect. \ref{sample}). This shows that the emission of the HCN  (0,\dh{0},\dh{0})  $J=2\rightarrow1$ line is distributed over a volume of the CSE at which its outflow 
has reached terminal velocity.

In some cases, the profile of the (0,\dh{0},\dh{0})  line has the shape expected from an angularly unresolved very optically thick line from an expanding CSE, a parabola with marked self-absorption of its low-velocity wing \citep[see, e.g.,][]{Olofsson1982,Morris1985}. %In most spectra 
%t is modified by self absorption in the line's low velocity portion, which 
%In most stars we observe deviations from a parabola and also
This self-absorption is also responsible for some of the differences between the values our data suggest for $\varv$$_{\inf}^{*}$ and/or $\varv$$_{\rm LSR}^{*}$ and published values. For example, for IRC+10216 (CW Leo), a somewhat higher value than our $-26.5~$\kms is frequently
given for the stellar velocity. %This is caused by self absorption in the low velocity wing of the HCN (and also the CO) $1-0$ line. %an effect that is not obvious in spectra with low SNR. 
Similarly, the literature velocity values might be biased for other stars, such as II Lup, RW LMi (CIT 6) and RAFGL 4211.
For CIT 6, the $\varv$$_{\rm LSR}^{*}$ and $\varv$$_{\inf}^{*}$ values we extracted from the (0,\dh{0},\dh{0})  line show the greatest
difference with published numbers. For this star and RAFGL 4211, our spectra led to an upward revision of $\varv$$_{\inf}^{*}$ compared to literature values. We note in particular that published spectra taken with a modest S/N can be affected by the bias described above. 

\begin{table*}
\caption{\label{results}Results of HCN $J=2\rightarrow1$ line observations}
%\begin{center}
%\setlength{\tabcolsep}{0.06cm} 
\centering
%`\begin{tabular}{lcclcccccccc}
%\begin{tabular}{llccllrr}
%\begin{tabular}{S[table-format=4.1]}{llccllrrcrr}
\begin{tabular}{lccrl}
\hline \hline

Object          & $(\nu_1,\nu_2,\nu_3)$& $\varv$$_{\rm LSR}$-range &\multicolumn{1}{c}{$\int T_{\rm MB}d\varv$}&\multicolumn{1}{c}{$R$} \\%\Mdot$&$\varv$$_\infty$
                                                                                 %        & $\varv$$_{\rm LSR}$ range 
                                                                                    %        &$\int T_{\rm MB}d$\\%$\varv$&$\Mdot$$^{\rm Mod}$&$r_{\rm out}$&$\theta_{\rm out}$&$X_{{\rm NH}_3}$\\

                    &                                    &(km~s$^{-1})$    &\multicolumn{1}{c}{(K km~s$^{-1})$} \\
                                                                                  
                                                                                   %&(\Mspy)                            
                                                                               %           &(km~s$^{-1}$   &(km~s$^{-1})$    &(K km~s$^{-1})$\\ &(\Mspy)&(cm) &($''$)\\
%\noalign{\smallskip\hline
%\noalign{\smallskip}
\hline
R For                            &(0,\dh{0},\dh{0})&[$-$18.5,+18.5]&5.8(1)&\multicolumn{1}{c}{--}\\ %0.70
                                     &\hml&[$-$6.1,+2.1]&0.59(5)&\dg9.8\\ %0.70
\vspace{0.07cm}
                                    &\hmld&--&$<0.16$&$> 36$\\ %0.70
R Lep                              &(0,\dh{0},\dh{0})&$[$-$4,+34.5]$&10.1(2)&\multicolumn{1}{c}{--}\\%0.47
                                      &\hml&$[+6.8,+16.5]$&1.02(8)&\dg9.9\\
\vs
                                       &\hmld        &--                    &$<0.23$&$>44$\\
AI Vol                            &(0,\dh{0},\dh{0})&[$-$54.4,-23.6]&15.7(1)&\multicolumn{1}{c}{--}\\%0.83 check vinf!
                                    &\hml&[$-$47.2,$-$31.3]&11.8(1)&\dg1.3\\
\vs
                                     &\hmld&[$-$42.2,$-$34.3]&1.14(9)&\dd14\\
X Cnc                             &(0,\dh{0},\dh{0})&[$-$19.6,-3.7]&1.20(5)&\multicolumn{1}{c}{--}\\%--CQ Pyx
                                     &\hml&--&$<0.14$&\dh$> 8.6$\\
\vs
                                    &\hmld&--&$<0.17$&\dh$>7.0$\\
CQ Pyx & (0,\dh{0},\dh{0})&[-11.3,+15.2]&16.4(1)&\multicolumn{1}{c}{--}\\%1.14
(RAFGL 4254)                                      &\hml&[$-$11.3,+11.3]&10.2(1)&\dg1.6\\
\vs
                                &\hmld&[$-$4.6,+3.1]&0.28(7)&\dd59\\
CW Leo    & (0,\dh{0},\dh{0})&[-47.0,-6.0]&615.4(2)&\multicolumn{1}{c}{--}\\%0.14
(IRC+10216)                               &\hml&[$-$40.9,$-$9.1]&99.2(2)&\dg6.2\\
\vs
                             &\hmld&[$-$39.7,$-$10.6]&9.7(1)&\dd63\\
X Vel                 &  (0,\dh{0},\dh{0})&[$\approx-$28.3,$\approx-$2.1]&2.25(9)&\multicolumn{1}{c}{--}\\%--
                        &\hml&$\approx-$26.2,$\approx-$16.9]&0.52(5)&\dg4.3\\
\vs
                &\hmld&--&$<0.25$&\dh$>9$\\
RW LMi  &  (0,\dh{0},\dh{0})&[-23.0,+21.6]&68.7(2)&\multicolumn{1}{c}{--}\\%0.46 check vinf!
(CIT 6)                  &\hml&[$-$23.0,+21.6]&36.4(1)&\dg1.9\\
\vs
                     &\hmld&[$-$16.4,+6.0]&1.47(8)&\dd47\\
U Hya               &   (0,\dh{0},\dh{0})&[$-$39.2,$-$22.3]&2.49(1)&\multicolumn{1}{c}{--}\\%--
                &\hml&--&$<0.30$&\dh$>8.3$\\
\vs
                &\hmld&--&$<0.29$&\dh$>8.6$\\\
V Hya                &(0,\dh{0},\dh{0})&[$\approx-$33,$\approx+7.1$]&9.3(2)&\multicolumn{1}{c}{--}\\%-- check vinf!
                &\hml&$\approx-$33,$\approx-$3]&6.2(1)&\dg1.5\\
\vs
              &\hmld&--&$<0.41$&\dh$>23$\\
V358 Lup & (0,\dh{0},\dh{0})&[$-$24.5,+22.1]&30.7(1)&\multicolumn{1}{c}{--}\\%0.95
(RAFGL 4211)                 &\hml&[$-$15.8,$+$6.7]&4.8(1)&\dg6.4\\
\vs
&\hmld&--&$<0.29$&$>106$\dh\\
X Tra          &(0,\dh{0},\dh{0})&[$-$15.8,+12.4]&2.56(1)&\multicolumn{1}{c}{--}\\%0.64
            &\hml&[$-$7.6,+7.8]&0.68(8)&\dg3.8\\
\vs
&\hmld&--&$<0.25$&\dh$>10$\\
II Lup &(0,\dh{0},\dh{0})&[$-$42.2,$-$15.2]&22.3(1)&\multicolumn{1}{c}{--}\\
                                                                                                &\hml&[$-$27.3,$-$3.2]&9.0(1)&\dg2.5\\
                                                                &\hmld&--&$<0.38$&$>59$\\

%\noalign{\smallskip
\hline
%\noalign{\smallskip}
\end{tabular}
%\end{center}
%\footnotesize{
\tablefoot{
Results of our APEX observations of the three $J=2\rightarrow1$ lines in three rows for each star (name in the first column): preceded by its vibrational quantum numbers, we list the LSR velocity range covered by emission and the velocity-integrated main-beam brightness temperature for each line, i.e., the line flux, for which the uncertainty in the last digit is given in parentheses. Upper limits ($3\sigma$) in this quantity for the \hmld\ line were calculated by assuming that it covered the same velocity range as the \hml\ line if detected, otherwise the range of the (0,\dh{0},\dh{0}) was assumed, as it was for upper limits in the \hml\ line. The rightmost column gives for the \hml\ and \hmld\ lines the ratio of these line fluxes to that of the (0,\dh{0},\dh{0}) line.%
%  mass-loss rate from the literature, terminal expansion velocity (from CO data), FWZP LSR velocity range 
%with observed emission, integrated main-beam brightness temperature, as well as, derived from our modeling,  mass-loss, outer radius of the NH$_3$ distribution (in cm and in arcseconds),  and   [\nhhh/H$_2$] abundance ratio. % For the $\varv$$_{\rm LSR}$ range and $\int T_{\rm MB}d$$\varv$  for each source, the first and second rows  give the values for the 
%\nhhh\ and the H$_2$O lines, respectively. %For both VY CMa and IRC+10420 we use the lower one of the two listed terminal velocities for modeling..
%The determination of the LSR ranges is somewhat subjective and the upper and lower velocities are uncertain by $\sim$ a few \kms\ for weaker lines.
%For all entries, the formal error in $\int T_{\rm MB}d$$\varv$ is smaller than 0.1 K~\kms.
%Numbers in braces relate to the references given below. For VY CMa we used the higher of the literature mass-loss rate values scaled to the recently measured trigonometric parallax distance; see \{3\}. For both IK Tau and IRC+10420 the lower and the higher values of $X_{{\rm NH}_3}$ are implied by the cm lines and the submm line, respectively.
%References for distances and mass loss rates are: 
%\tablebib{
%\{1\} \citet{Hale1997};
%\{2\} \citet{Decin2010}; 
%\{3\} \citet{Choi2008};l
%\{4\} \citet{Decin2006}; 
%\{5\} \citet{Justtanont2006}; 
%\{6\} \citet{Kemper2003}
%\{7\} \citet{trung2009}.
}
\end{table*}

The \hml\ maser line is detected   in 11 sources. Toward some of them, its flux rivals that of the (0,\dh{0},\dh{0})  line. In contrast, emission in the \hmld\ line is  only found toward  four of the sources with \hml\ emission and is always much weaker. 

We do not see a propensity for a detection of the maser line in sources with higher mass-loss rates or a clear relation between  $R$ and the mass-loss rates of the objects, but we note that the maser line remains undetected toward X Cnc and U Hya and is barely detected toward X Tra, three sources with the lowest mass-loss rates in our sample.

In all cases, the \hml\ maser line profiles are asymmetric. In several objects this line shows a single narrow feature within a few \kms\ of the stellar velocity that is superposed on broader emission. Except for the cases of CW Leo and V Hya, which we discuss separately (in Sections \ref{cwprofile} and \ref{others}), the \hml\ line covers a significantly narrower velocity range than the (0,\dh{0},\dh{0})  line and in AI Vol, CQ Pyx, and RW LMi the \hmld\ line covers an even narrower range. For most sources the bulk of emission in the \hml\ maser line is blueshifted relative to the stellar velocity. The possible significance of this is discussed in Sect. \ref{region}.

%\subsection{Line profiles}
%\subsection{Line velocities and profiles}
%Stellar velocities, $ and terminal velocities, $\vinf}, for all stars in our sample are listed iisted in the catalog by \cite{Loup1993{

%In the inner parts of CSEs rotational lines from a variety of vibrational states, including overtones of the bending mode (020--080) mixed ones (e.g., 111 or 011)  \citep{Cerni1996, Cerni2011}, even strong laser emission in the 

\section{\label{discussion}Discussion}
The vibrationally excited lines cover a narrower velocity range than the line from the vibrational ground state. The reason for this most likely is that they arise from 
a hot region of the CSE close to the stellar surface in which dust is still forming and the outflow has not yet reached its terminal velocity. Meaningful information on this region for a C-rich AGB star is so 
far only available for IRC+10216 from an analysis of many HCN transitions; see Sect. \ref{hotcw}, in which a distance of 40 stellar radii
is estimated for the size of this region, or $\approx  0\as5$. Long-baseline ALMA observations will resolve this region, as they have done in the case of the O-rich (M-type) AGB star $o$ Ceti for emission from the SiO and \hzo\ molecules \citep{Wong2016}.

By far the most extensive information on excited HCN in the literature exists on IRC+10216, and  we now concentrate on this star. 

\subsection{\label{cwleo}IRC+10216}
\subsubsection{\label{cwprofile}Variability: Comparison of the
IRC+10216 line profiles from two epochs}
\begin{figure}
%\centerline{\resizebox{0.75\hsize}{!}{\includegraphics[angle=0]{fig2.eps}}}
%\centerline{\resizebox{0.65\hsize}{!}{\includegraphics[angle=0]{fig3.eps}}}
\centerline{\resizebox{0.75\hsize}{!}{\includegraphics[angle=0]{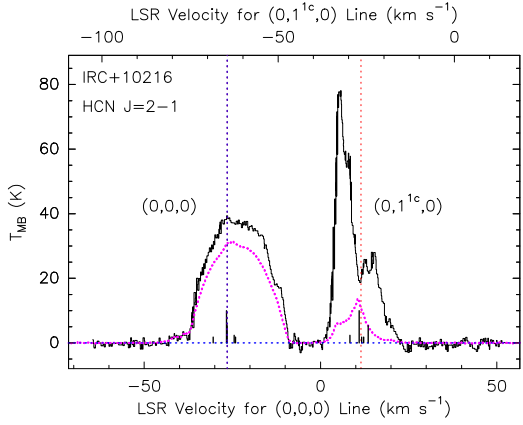}}}
\caption{Comparison of the spectra for the HCN $J=2\rightarrow1$ (0,\dh{0},\dh{0})  and \hml\ lines taken with the IRAM 30 m telescope in 1989 April by \citet{LucasCernicharo1989} (full line spectrum)
and with the APEX 12 m telescope on 2015 May 28 (dotted magenta line spectrum). The bottom and top LSR velocity scales are appropriate for the (0,\dh{0},\dh{0})  and the \hml\ lines, respectively.
Both spectra have a velocity resolution of 0.26. \kms The vertical bars give the relative intensities and velocities or the hfs components of the two lines. The intensity of the strongest component is normalized to $T_{\rm MB}$ = 10 K.}
\label{comp}
\end{figure}

Figure \ref{comp} shows spectra of the HCN $J=2\rightarrow1$ (0,\dh{0},\dh{0})  and \hml\ lines taken 26 years apart. The top and bottom spectra were taken with  the IRAM 30 m and the APEX 12 m telescopes, 
respectively. The IRAM spectrum was extracted from Fig. 1 of \citet{LucasCernicharo1989}\footnote{We used the WebPlotDigitizer tool available on \url{http://arohatgi.info/}}.  The intensity of the (0,0,0 line) is comparable in the two spectra, 38 vs. 31 K. Since the emission from this line arises from an extended region of the CSE with a size of thousands of au (or 
stellar radii) and is determined by thermal processes (see below), we expect little variability of its profile. We can estimate the size of the emission region: Assuming a 
Gaussian telescope beam with an FWHM size $\theta_{\rm B(A~or~I)}$ for both the APEX and the IRAM  telescopes (superscripts A and I) 
%\end{document}
and a (simplifying) Gaussian source size of the HCN $J=2\rightarrow1$ (0,\dh{0},\dh{0})  line emission region with an FWHM size $\theta_{\rm S}$, the observed main-beam brightness temperature is given by 
\begin{equation}
%T^{\rm A~or~I}_{\rm MB} = { \theta^2_{\rm S} \over {\theta^2_{\rm S} + \theta^2_{\rm B}{\rm (A~or~I)}} }T_{\rm B},
T^{\rm A~or~I}_{\rm MB} = { \theta^2_{\rm S} \over {\theta^2_{\rm S} + \theta^2_{\rm B(A~or~I)}} }T_{\rm B},
\label{Eq1}
\end{equation}
%\end{document}
where $\theta_{\rm B(A)} = 36''$ and  $\theta_{\rm B(I)} = 15''$. $T_{\rm B}$ is the brightness temperature of the emission region. Dividing this
expression for the IRAM 30 m telescope by that for the APEX telescope, we can solve for the source size  $\theta_{\rm S}$, which is given by 
\begin{equation}
\theta_{\rm S}  = \bigg({{ \theta^2_{\rm B(A)} - R_{\rm IA}~\theta^2_{\rm B(I)} }\over {R_{\rm IA}-1}}\bigg)^{1/2}.
\label{Eq2}
\end{equation}

Here $R_{\rm IA}$ is the ratio of the main-beam brightness temperatures measured in the IRAM to that in the APEX beam, that is, 36 K/31 K = 1.16. Plugging in numbers, Eq. \ref{Eq2} gives a size of $80''$ for the  region from which the emission of the HCN $J=2\rightarrow1$ (0,\dh{0},\dh{0})  line arises and a strict lower limit of $48''$
when we assume an uncertainty in  the $T_{\rm B}$ scale of 10\%.
The nominal value of $80''$  is higher than expected, given that the emission region of the lower-critical density HCN $J=1\rightarrow0$ line has been measured to have an extent of $64''$ by \citep{DayalBieging1995} with the Berkeley-Illinois-Maryland Array (BIMA). Given the calibration uncertainties in typical datasets obtained with (sub)millimeter single-dish telescopes, it is impossible to obtain a reliable estimate of the size of the $J=2\rightarrow1$ (0,\dh{0},\dh{0})  line emission region without mapping. % it is difficult to estimate the size of the $J=2\rightarrow1$ (0,\dh{0},\dh{0})  line's emission region. 

Regardless of the uncertainties in the absolute intensity calibration, when we inspect the spectrum of the (0,\dh{0},\dh{0})  line closely (Fig. \ref{comp}), we find other  possible evidence for variability, namely the ``shoulder'' in its shape near an LSR velocity of $-12$~\kms, which is observed in the spectrum taken with the IRAM telescope, but not in the APEX spectrum. 
%Apart from calibration uncertainties, 
%Actual variability in this line might affect its observed appearance. 
Variability of non-maser spectral lines might be expected and has been observed in high-excitation lines arising in the innermost 
CSE of IRC+10216 \citep{Cerni2014, He2017}. However, variability has recently also been reported for lines from some molecules whose emission arises in the outer envelope of the star, while lines from others, for example, from SiC$_2$, remain constant \citep{Cerni2014, He2017}. Even the very large scale  continuum emission from dust  has been found to show variability, which 
is explained as a reflex to the stellar intrinsic variability \citep{Groenewegen2012}. 
%A caveat: While, on face value, variability in the (0,\dh{0},\dh{0})  line seems evident from the different peak intensities in the two spectra. %: 38 K in the IRAM spectrum vs. 32 K in the APEX spectrum (KARL: VERIFY EXACT NUMBERS). 
%While such a difference is well within the typical absolute calibration uncertainty of mm-wavelength single dish telescope datasets, we note that increasing the intensity of the line measured with APEX 12 m telescope relative to that measured with the IRAM 30 telescope would imply an even larger source size than that derived above. 

While the (0,\dh{0},\dh{0})  line may show some variability, in contrast, it is highly obvious from the very different shapes of the two spectra of the \hml\ line alone that its profile changed dramatically over the 
years. Thus, using Eq. \ref{Eq2} to determine the size of its emission region, which is certainly very compact, does not make sense. For a point-like source ($\theta_{\rm S}\rightarrow 0$), $R_{\rm IA}$ is just given by the 
squared ratio of the IRAM and the APEX FWHM values, that is, 5.8. This ratio is indeed approached for some portion of the \hml\ line spectrum. %, i.e., between ... and ... \kms. 
However, in particular at velocities lower than the systemic velocity, that is, $-26.5$  \kms\ (upper velocity scale in Fig. \ref{comp}), dramatically stronger emission has been measured by the IRAM 30-meter telescope than by the APEX telescope  26 years later.
We revisit the $\nu_2=1$ line variability in Sect. \ref{region}.

%\end{document}

\subsubsection{\label{hotcw}Hot HCN around IRC+10216}

To place our results into context, we note that  \citet{Cerni2011} studied a total of 63 $J=3\rightarrow2$ rotational transitions from 28 
vibrational states with energies of up to 
 10700 K toward IRC+10216. Assuming local thermodynamic equilibrium (LTE) and using a rotation diagram (``Boltzmann plot'') analysis (their Fig. 2), these authors find the lines to arise from three  
 temperature regimes, characterized  by vibrational temperatures
\Tvib of $\approx 2465$, 1240, and 410 K. The 
first of these  is close to the effective temperature of 2750 K of the star inferred by \citet{Menten2012}, and, together with the 
also determined column density, implies that HCN exists close to the stellar surface with an abundance relative to 
 molecular hydrogen, $x$(HCN), of 5--7$\cdot10^{-5}$. The extremely high excitation ($\sim4500$~K) laser lines discussed in Sect. \ref{intro} have widths %($\approx 5~$\km_ 
 consistent with an origin in this  zone. Recent ALMA observations of lines from vibrationally excited HNC (hydrogen \textit{iso}cyanide) suggest that these 
 exclusively originate from this innermost region \citep{Cerni2013}.
 
Over the second to the third zones (at 1240 and 420 K), which reside successively farther away from the star, $x$(HCN) drops by an order of magnitude, and the FWHM line widths increase from the near-photospheric value of 5 \kms\ to 19 \kms\ in zone III. From visual inspection of spectra presented by \citet{Cerni2011}, we find the FWZP of zone III lines to be $\approx 30$~\kms, identical to the value we find for our $\nu_2=1$ lines.
The lines fitted by Cernicharo et al. (2011) in their Boltzmann plot, whose slopes yield the rotational temperatures or the three zones, represent all the observed data points fairly well. This shows that LTE is a good assumption for the populations of non-inverted, that is, of most, HCN levels throughout the innermost CSE of
IRC+10216. 

\citet{Tenenbaum2010} and \citet{Patel2011}, for the HCN $J=3\raw2$ and $4\raw$ rotational lines confirm this dichotomy of narrow lines from highly excited vibrational states and broader lines from lower excitation states and the ground state.

Remarkably, both the strong $(0,1^{1_{\rm c}},0)$ maser \textit{and} the weak $(0,1^{1_{\rm 
d}},0)$ line cover exactly the velocity range \citet{Cerni2011} determined for the zone III 
lines. Such broad widths, which correspond to twice the CSE terminal velocity, have been 
reported for vibrationally excited HCN lines before. The interpretation
was that they  arise from 
the entire region in which dust grains nucleate, and that consequently, the gas in the stellar 
outflow is accelerated from the stellar surface out to 20 \rstar \citep{Fonfria2008, 
Cerni2011, Cerni2013}. This corresponds to 38 au, using the value of 1.9 au that \citet{Menten2012} have derived for \rstar,
the %\LEt{again, please check whether something is missing here}} 
\textit{radius} of 
the stellar optical photosphere. These authors directly measured a diameter of 10.8 au (83 milliarcseconds) for the radio (and mm) photosphere of IRC+10216, which is significantly larger than the optical photosphere..

%In contrast, the emission in the HCN laser lines that arise from very high excitation state (see Sect \ref{intro}) shows a single narrow feature centered on the stellar velocity that has a width of $\approx 5~$\kms, matching that of the 

\subsection{\label{others}Other sources}
As discussed in Sect. \ref{results}, for most sources, the \hml\ line covers a much narrower velocity range than the (0,\dh{0},\dh{0})  line and shows a variety of shapes. We now comment on this and other properties of selected sources. 

In some sources the profile is dominated by one very narrow component, perhaps most clearly manifested in R For, which at our spectral resolution only appears in a few channels. This feature is most typical for maser lines, and it alone can be taken as indication for maser action. In addition to this narrow component, most sources show an underlying broader component that can have the form of a relatively regular pedestal, like that in AI Vol, or an irregular shape. % that looks like a blend of several narrower components. 
This broader component might represent  thermal emission, but because of its intensity, in particular compared to that of the \hmld\ line, it is more likely a blend of several weaker maser spikes. Such features, which are also observed in other masing lines, may be highly variable and significantly change the overall appearance of the line profile with time. Except for V Hya (see Sect. \ref{vhya}), the velocity range covered by the broader component is much smaller than the full width of the (0,0,0) HCN line.

\subsubsection{\label{cit6}CIT 6 (RW LMi)}
In some respects, CIT 6 appears to be a twin of IRC+10216. This pertains to their similar periods and terminal velocities, although the mass-loss rate of the latter object has been estimated to exceed the rate of CIT 6 by a large significant factor. Nevertheless, if we scale the line fluxes of CIT 6 to those of IRC+10216 with the square of their distances, we obtain comparable values. In contrast to IRC+10216,  the vibrationally excited  lines cover  a narrower velocity range toward CIT 6 than the (0,\dh{0},\dh{0})  line. 

We note that for CIT 6, the \hml\ line has a higher peak brightness than the (0,\dh{0},\dh{0}). The same is true for AI Vol (and for CQ Pyx both lines are comparable).
In stark contrast, %remarkably, 
for another HCN maser line from an excited bending-mode state, much stronger emission was found from CIT 6 than from IRC+10216: \cite{Guilloteau1987} found the emission in the $(0,2^0,0$) $J=1\rightarrow0$ line, the first HCN maser detected, to be $\approx50$ times stronger than in IRC+10216; see also Sect. \ref{variability}.

\subsubsection{\label{vhya}V Hydrae}
V Hya is a  very evolved C-rich object that has been termed a ``dying star''. It appears to be in transitory state to a planetary nebula \citep{Tsuji1988, Kahane1988, Sahai2003}.  %It has developed high velocity outflow that has been 
Observations of  lines observed at optical, IR, and millimeter
wavelengths present a complex picture that has been interpreted as showing a high-velocity bipolar outflow \citep{Kahane1996} together with an equatorial structure perpendicular to it \citep{Sahai2003, HIrano2004}.

 As far as we know, our result for
 the HCN \hml\ line represents the first detection of (sub)millimeter-wavelength rotational emission toward V Hya 
from highly excited energy levels. % within an excited  rotational state toward V Hya. 
Quite remarkably,  we measure a flux in %the 
this line that is comparable to the flux in the (0,\dh{0},\dh{0})  line. 
 Moreover, both  lines cover comparable velocity ranges and have even similar profiles as the HCN $J = 
 1\rightarrow0$ line observed with the Nobeyama 45 m telescope \citep{Tsuji1988}.  These velocity ranges imply that this emission from all these lines arises from much of the acceleration region of this object's outflow.
 
%The The reason for the high width of the base of the (0,1^1c,0) transition relative to that of (0,0,0) in V %Hya is not understood but may be related to the location of this Mira in a triple system (so that the %outer envelope of this object is truncated) or unusually high rotation velocity (Barnbaum et al. 1995).

\subsubsection{\label{4211}Other stars}
\citet{Smith2014} summarized observations of maser emission in the $(0,2^0,0)$ $J = 1\rightarrow0$ line, which before had been detected
in a total of nine sources.  To this list they added V358 Lup ($\equiv$ IRAS 15082$-$4808, RAFGL 4211). Three of these objects
are part of our sample, namely IRC+10216, CIT 6, and V358 Lup.  Toward the last source, maser emission had been observed to have a different appearance in 2010 and 2011 and was not detected at all in 1993. In 2011, a single feature was observed with a flux density of $\approx 4$~Jy at a velocity that was a few \kms\ blueshifted from the stellar systemic velocity.  In 2010, the intensity of this feature had been higher,  $\approx 15$~Jy, while lower-intensity emission had appeared at higher velocity. The total velocity range covered by this is smaller than that we observe in the $J = 2\rightarrow1$ \hml\ line, which has about twice the peak intensity of the $(0,2^0,0)$ line. %Neither for V358 Lup, nor for CIT 6, for which also data for multiple epochs exists, a propensity for stronger maser emission near $\phi_{\rm IR} = 0$ or weaker maser emission near $\phi_{\rm IR} = 0.5$ is observed.
For V358 Lup and CIT 6, data also exist for multiple epochs. Neither source shows a propensity for stronger maser emission near $\phi_{\rm IR} = 0$ or weaker maser emission near $\phi_{\rm IR} = 0.5$.

\subsection{Properties of the \hml\ maser emission}
\subsubsection{\label{pumping}Maser pumping}
Based on observations of (non-masing) $J = 2 -1 $ (and other) rotational HCN lines from within several different vibrational states, Lucas \&\ Cernicharo suggested that 
the \hml\ 
maser could be explained by pumping into the \textit{v}$_2 = 1$  and 2 states via IR radiation at 14 and  $7~\mu$m, respectively. For radiative pumping to be 
feasible, the availability of at least one pump photon per maser photon is required. In Table \ref{lumi} we compare the 
photon 
luminosities, $L_{\rm ph}({\rm M})$,  in that line with the continuum photon luminosities in the 12 $\mu$m IRAS band, $L_{\rm ph}(12 \mu{\rm m}), $ for the 11
sources with detections in the \hml\ maser line. Values for the latter were 
calculated from the 12 $\mu$m flux 
densities listed in Table \ref{sample}, scaled by the velocity range over which maser emission is observed, taken from Table \ref{results}. How does the 12 $\mu$m flux density 
compare to the 7 and 14 $\mu$m flux densities? Detailed modeling of IRC+10216's wide band spectral energy distribution (SED),
presented in the appendix of \citet{Menten2006}, shows that it peaks between $\approx 8~{\rm and}~10~\mu$m (depending on its phase) and that 
its 7 and  $14~\mu$m flux densities are within a factor of 2 of the 12 $\mu$m flux density. Moreover, all these flux densities vary by less than a factor of 5 over the 
stellar pulsation  cycle. Since IRC+10216's $L_{\rm ph}(12 \mu{\rm m})$ of $1.8\cdot10^{46}~{\rm s^{-1}}$ is greater by far than its $L_{\rm ph}({\rm M})$ of $3.9\cdot10^{42}~{\rm s^{-1}}$, it is safe to say that 
radiative pumping is certainly feasible in principle, even if the uncertainties discussed above are taken into account. This is also very likely true for the other stars in our sample,  although detailed SED modeling is not available for them;  in all cases, $L_{\rm ph}(12 \mu{\rm m}) \gg L_{\rm ph}({\rm M})$.

We would like to remark that the photon luminosities we determine for the HCN $J = 2 -1$ \hml\ maser line (Table \ref{lumi}) are comparable to or higher than typical values found for the strongest ($J = 1 - 0$ or $2 -1$)  vibrationally excited ($\varv=1$ or 2) SiO maser lines toward O-rich AGB stars.

\subsubsection{\label{variability}Relation between maser flux and IR light curve}
The flux of SiO masers around O-rich AGB stars shows a very close correlation with the IR light curve of these O-rich stars. The maser flux attains its maximum at visual phase between $\approx 0.05$ and 0.2, that is, at or near the maximum of the IR light curve \citep[see, e.g., ][]{Bujarrabal1987, Pardo2004}. 

Since only one of the \hml\ masers has been observed twice, we can only make very qualitative statements on the relation of its flux to the continuum flux from the star or its dust shell.
The epoch when \citet{LucasCernicharo1989} first detected the \hml\ and \hmld\ lines toward IRC+10216, 1989 April (Julian day, JD $2447631\pm15$), falls within the time range over which \citet{LeBertre1992} monitored the brightness of this star in multiple IR bands. Inspecting his published $J, K, L,$ and $M$ light curves, we estimate that the star was  at IR phase, $\phi_{\rm IR}$ =0.23. %(TOMEK, CHECK!). 
For our own observations, which were made on JD 2457171, using the JD of IR maximum and the value for the period determined by \citet{Menten2012}, we estimate $\phi_{\rm IR} \approx 0.23$, an identical value. Given that the maser luminosity was much stronger at the first epoch, this is very surprising.

\begin{table}
\caption{\label{lumi}Comparison of HCN maser and IR photon luminosities}
%\begin{center}
%\setlength{\tabcolsep}{0.06cm} 
\centering
%`\begin{tabular}{lcclcccccccc}
%\begin{tabular}{llccllrr}
%\begin{tabular}{S[table-format=4.1]}{llccllrrcrr}
\begin{tabular}{lrrr}
\hline \hline

Object          &$S_{12\mu{\rm m}}$& $L_{\rm ph}(12 \mu{\rm m})$& $L_{\rm ph}({\rm M})$ \\
                    & (Jy) &$10^{42}~{\rm  s}^{-1}$    & $10^{42}~{\rm  s}^{-1}$ \\
                                                                                    
                                                                                   %&(\Mspy)                            
                                                                               %           &(km~s$^{-1}$   &(km~s$^{-1})$    &(K km~s$^{-1})$\\ &(\Mspy)&(cm) &($''$)\\
%\noalign{\smallskip\hline
%\noalign{\smallskip}
\hline
R For  & 254& 620 &6\\
R Lep  & 380& 1000&5\\
AI Vol  & 613 & 4100&200\\
X Cnc  & 90 & 400&$<1.3$\\
CQ Pyx & 737 & 13000&300\\
CW Leo & 47500& 18000&39\\
X Vel     & 89 & 200&5\\
RW LMi & 3320 &19000&160\\
U Hya    &206& 260&$<0.7$\\
V Hya    &1110 & 2200&14\\
V358 Lup & 793 & 9700&87\\
X Tra    & 201 & 410&3\\
II Lup & 1320 & 4200&40\\
%\noalign{\smallskip
\hline
%\noalign{\smallskip}
\end{tabular}
%\end{center}
%\footnotesize{
\tablefoot{
Comparison of IR and \hml\ maser luminosities. Columns are (left to right) the name, flux density in the 12 $\mu$m IRAS band (taken from the IRAS Point Source Catalog \tablefootnote{\url{https://heasarc.gsfc.nasa.gov/W3Browse/all/iraspsc.html}}), and the photon rates emitted in the maser line and at $12~\mu m$ in a wavelength range that corresponds to the velocity range covered by maser emission (taken from Table \ref{results}). In the two cases for which no maser line was detected (X Cnc and U Hya), upper limits were calculated using the velocity range covered by emission in the (0,\dh{0},\dh{0})  line and the $3\sigma$ upper limits for the maser-line integrated flux density given in that table.}
\end{table}

A ``decoupling'' of IR light curve and maser variability has also been found for the $(0,2^0,0)$ $J = 1\rightarrow0$ maser line observed toward  IRAS 15082$-$4808 (RAFGL 4211) by \citet{Smith2014}. For the few C-rich stars for which multiple observations of this line exist, these authors could not establish a propensity of stronger maser flux to appear at the IR maximum. The weakness of this line when it was first  detected in IRC+10216 has been mentioned in Sect. \ref{cit6}. %At the date of those observations, 1986 December 6 (JD 2446770.5), we read off  $\phi_{\rm IR} \approx 0.9$ from IRC+10216's  $J, K,$ \&\ and $M$ light curves  \citep{LeBertre1992}. (TOMEK, CHECK!) 
Multiple  observations at different epochs have always found it at similarly weak ($\approx 2$~Jy) intensities and with similar (simple) line profiles \citep{Lucas1986, Lucas1988, Guilloteau1987, LucasGuilloteau1992}. This makes it doubtful whether this line shows maser action toward IRC+10216.

To explore this issue further, we note that other HCN lines from various vibrationally excited states show variability in IRC+10216
as well, but at a much lower level than the $J=2\rightarrow1$ \hmld\ line \citep[see Sect. \ref{cwprofile} 
and ][]{Cerni2014, He2017}. \citet{Cerni2014} observed the HCN   $J=6\rightarrow5$ \hml\ and (0,\dh{0},\dh{0})  lines with the Heterodyne Instrument for the Far Infrared (HIFI) on board the Herschel Space Observatory at 
two different times, on 2010 May and 2010 November, when the phase of the star was 0.23 and 0.53, respectively. On both dates, the  $J=6\rightarrow5$ (0,\dh{0},\dh{0}) line shape was very similar to that of the 
$J=2\rightarrow1$  (0,\dh{0},\dh{0})  line observed by us (see Sect. \ref{cwprofile}), while its total intensity decreased by 10\% between the first to the second date. Changes in the  $J=6\rightarrow5$ \hml\ line were 
much more dramatic.  On both dates, this line showed a sloping profile over most of the velocity
range, which decreased in intensity from lower to higher velocities, in contrast to the profile we observed for the  $J = 2\raw1$ \hml\ line, which peaks at the systemic velocity. Between 2010 May and November, the intensity of this broad emission decreased by a factor of $\approx 1.6$ in total. On the two 2010 dates, a narrow 
spike is observed on the extreme blue edge of the line, whose excess intensity (over that of the the broad emission) decreased in intensity even by a larger factor or $\approx 2$. The intensity of a  narrow 
spike  at the redshifted edge diminished by an even larger amount. We note that the  \hml\ $J = 2\raw1$ line also had pronounced blueshifted emission when \citet{LucasCernicharo1989} 
observed it in 1989. 

\cite{He2017} discussed the variability in the \hmld\ $J=3\raw2$ line %\footnote{\citet{He2017} use variant nomenclature for the vibrationally excited state, i.e, $1^{1{\rm f}}$.} 
over the stellar 
(IR) cycle and considered different velocity ranges separately. While they found that the red portion of the profile remained constant, they discussed pronounced changes in its blue part that appeared to be
correlated with the IR light. As one possibility to explain this, they suggested that the photospheric continuum emission might
be amplified, which would require that the HCN maser were unsaturated. We 
note that the  \hml\  $J= 2\raw1$ line shows a very different profile at the two times it was observed, when the star was at \textit{\textup{identical}} phase and that the  \hml\ $J= 6\raw5$ line profile was different from both while also observed at this very same phase. This does not support background 
amplification, which, as discussed in Sect. \ref{region}, is generally rarely found to be associated with maser emission in AGB stars (see Sect. \ref{region}).

In conclusion, the flux of  vibrationally excited HCN maser lines does not appear to be correlated with the stellar cycle, in contrast to the case of SiO masers around O-rich  AGB stars. The reason for this is unclear, but we point out that the C-rich stars have mass-loss rates higher
by more than an order of magnitude than O-rich AGB stars. This causes C-rich stars to have higher densities close to their photosphere than O-rich objects, possibly increasing the relative importance of collisional pumping for the former. In this context, it is interesting to note that \citet{Pardo2004} did not find any correlation between SiO maser and stellar  phase for O-rich red supergiants, which have yet higher mass-loss rates than C-rich AGB stars. 

In the case of SiO, the correlation of maser flux to stellar flux is frequently taken as evidence for radiative, rather than collisional pumping to cause the inversion \citep[see, e.g., ][]{Pardo2004}; both schemes have been discussed in the literature \citep{Lockett1992}. Our results seem to indicate the possibility of a collisional pump for the vibrationally excited HCN masers discussed here. However, we point out that as discussed in Sect. \ref{pumping}, there are abundant IR photons available to allow radiative pumping at any phase of the stellar variability cycle.

\subsubsection{\label{linewidths}Maser line widths}

Under the assumption that the line width of narrow maser spikes in our observed spectra is 
determined by thermal broadening, we can derive a minimum temperature for the masing 
region. The kinetic temperature, $T_{\rm kin}$, causing the  thermal broadening of the profile of an HCN 
line to an FWHM $\Delta \varv,$ is given by 
\begin{equation}
T_{\rm kin} = (8~ln 2~k)^{-1}  \Delta \varv^2 m \approx 586~\Bigg({{\Delta v}\over{\kms}}\Bigg)^2 ~{\rm K},  %\sqrt{8 ln 2 k T/m} \approx 0.215 \sqrt{T/m}
\end{equation}
where $k$ is the Boltzmann constant, and the mass, $m$,  of an HCN molecule is 27 atomic mass units. The 
narrowest features in our spectra have $\Delta \varv \approx $ 0.6--1.2~\kms when we only 
determine the width of the narrow components that are superposed on broader emission; see Figs. 
\ref{four} and \ref{rest}. This corresponds to T = 215--844 K, which is broadly consistent with 410 K, the value 
\citet{Cerni2011} determine for zone III in IRC+10216 (see Sect. \ref{hotcw}). We note that the above only provides 
a very qualitative estimate of the kinetic temperature and does not account for possible line 
narrowing that could occur for unsaturated maser lines. However, together with the fact that 
turbulence would broaden the lines, this makes it a lower limit of the actual temperature in the 
maser-emitting region.

As a consistency check, we note that using the Stefan-Boltzmann law and the luminosity of  8640 \Lsun\ determined by \citet{Menten2012}  for IRC+10216, we derive a temperature of 614 K for  a region of radius 38 au around this star. This is again comparable to the 410 K derived by \citet{Cerni2011}, taking into account its variability.

We exclude line broadening caused by hyperfine structure. To illustrate this, the second strongest hfs component of the \hml\ line has a an intrinsic intensity of 0.54 of that of the 
strongest and a velocity offset of +2.53 \kms\ relative to it. In sources with narrow features ($\Delta \varv \simless 1$~\kms), 
for example, R For or RAFGL 4211 (Fig. \ref{rest}), we do not find any feature at this velocity offset. This confirms 
the high-gain maser nature of this line emission: Exponential amplification of unsaturated maser emission with substantial gains very strongly favors the strongest hfs component and leaves intrinsically lower intensity components very weak and even undetectable
in our case. 

\subsubsection{\label{region}Constraints on the maser emission}
As mentioned in Sect. \ref{results}, the peak emission in the \hml\ maser line is slightly blueshifted relative to the systemic velocity for several of our sources, as is the $(0,2^0,0)$ $J = 1\rightarrow0$ maser line in RAFGL 4211; see Sect. \ref{4211}. This effect would be naturally expected if the masing material were expanding 
away from the star and it would amplify the millimeter-wavelength continuum from the stellar photosphere. In reality, however, such amplification of photospheric continuum emission is rarely % , if ever, %(if at all) 
observed for circumstellar masers. %\footnote{
It has recently been invoked by \citet{Gong2017} to explain the fact the blueshifted emission is variable and stronger than the redshifted emission for the $J=1\raw0$ line of SiS, whose maser nature they prove for IRC+10216.  

High angular resolution observations of vibrationally excited SiO masers around O-rich AGB stars invariably show rings with 
sizes of a few \rstar\  around the stars, indicating tangential amplification \citep[see, e.g., ][]{Cotton2004,ReidMenten2007,Gray2009}. Even more extreme, in the case of the 
archetypical M-type AGB star $o$ Ceti (Mira), ALMA observations have recently shown \textit{redshifted absorption} in vibrationally excited SiO lines (and one H$_2$O line) toward the photosphere of the star, surrounded by rings of maser emission \citep{Wong2016}.  By analogy, we conclude that simple inferences merely based on spectra are to be taken with caution for our HCN data, and high-resolution imaging with ALMA is 
required for an understanding of the complex up to a few tens of \rstar-sized regions around C-rich AGB stars from which vibrationally excited HCN emission arises. So 
far, the physical conditions and (indirectly) the size of this  region is only characterized by multiple high-excitation line spectroscopy (and not yet interferometry) and only for IRC+10216; see Sect. 
\ref{hotcw}. 

In Sect. \ref{hotcw} we have argued that for IRC+10216, the diameter of the emission region for both the strong \hml\ maser and the %non-masing 
\hmld\ line is the 76 au inferred by \citet{Cerni2013} from their modeling of HCN lines from various vibrational excited states. This corresponds to an angular size $\theta_{\rm S} = 
%T^{\rm A~or~I}_{\rm MB} = { \theta^2_{\rm S} \over {\theta^2_{\rm S} + \theta^2_{\rm B(A~or~I)}} }T_{\rm B},
0\as63$ at the distance of this star. Taking this together with the peak main-beam brightness temperature of $T^{\rm A}_{\rm MB}$ = 14 K observed with the APEX 
telescope (see Fig. \ref{four}), we can use Eq. \ref{Eq2} to calculate a brightness temperature of $T_{\rm B} =  46000$~ K for the \hml\ maser line. For this line, measured with the IRAM 30  m telescope by \citet[][see also our Fig. \ref{comp}]{LucasCernicharo1989}, we calculate $T_{\rm B} =  43000$~ K. These represent lower limits as the strongest maser emission may arise from a smaller region than assumed here. These 
values are higher than any value that would be consistent with thermal excitation, proving the maser nature of this line. 

For the \hmld\ line ($T^{\rm A}_{\rm MB} = 
0.8$ K), we obtain a lower value of  $T_{\rm B} =  2600$~K. Even this is significantly higher than the $\approx 400$~K one would expect for an optically thick line from zone III (see Sect. \ref{hotcw}). 
It likely implies weaker maser action in this line as well. Comparing the spectrum observed by us for this line with that observed by \citet{LucasCernicharo1989}, we see marked changes, both in shape and intensity. In contrast to the strong \hml\ maser line, we did not find dramatically stronger emission for the \hmld\ line in 2015 at 
velocities lower than the systemic value than in 1989, and despite the relatively low S/N of this line observed in 1989 with the IRAM telescope, we find a different profile with APEX. Moreover, for a compact source of size $0\as63,$ one would expect $T^{\rm I}_{\rm MB} /T^{\rm A}_{\rm MB} = 5.8$, whereas we observe a value $\approx 1.5$. This suggests significant variability in the \hmld\ line as well, supporting possible maser action. 
%To probe the inner parts of IRC+10216's CSE, \citet{Cerni1999} used the Short Wavelength Spectrometer (SWS) aboard the Infrared Space Observatory (ISO) to study direct ro-vibrational lines from the stretching and the bending modes, including overtone and combinations bands. Toward the same star, pure rotational lines from a variety of vibrational states, including overtones of the bending mode (0,2,0--0,8,0) and combination bands (e.g., 1,1,1 or 0,1,1)  were observed with the ISO Long Wavelength Spectrometer (LWS)  with upper levels J's, $J_{\rm u}$, of 18--48 \citep{Cerni1996} and with the IRAM 30 meter telescope ($J=3-2$, \citep{Cerni2011}).
%, even strong laser emission in the 

%\subsubsection{$l$-type doubling and the $\nu_2=1, J=2-1$ maser line}
%The main focus of this paper is on the maser emission in 

%\subsection{\label{modeling}}
%\footnotesize{
%\bibliographystyle{aa}
%\bibliography{KMM}

%\end{document}

%\end{document}
%}

\section{\label{summary}Summary and outlook}
Using the new ALMA Band 5 prototype receiver on the APEX 12-meter telescope, we conducted a survey for emission in the $J = 2\rightarrow1$ rotational line from the (0,\dh{0},\dh{0}) , \hml,\ and \hmld\ vibrationally excited states of HCN  toward a sample of 13 carbon-rich AGB stars. We detect broad thermally excited emission in the (0,\dh{0},\dh{0})  line toward all of them and strong maser emission  in the \hml\ line toward a total of 11. Toward 4 of the latter, we also detect much weaker emission in the \hmld\ line. The velocity ranges covered by the vibrationally excited lines are consistent with their origin from within several stellar radii of the stellar photospheres, that is, in regions where the dust is still forming and the outflow is accelerating.

While it is clear that abundant IR photons are available to allow radiative pumping of the \hml\ maser line, limited information on this and another vibrationally excited HCN maser line indicate a collisional pumping process. Temporally extended time monitoring of a sample of objects with the APEX telescope will shed light on the issue.

ALMA has now been fully equipped with Band 5 receivers and will allow interesting high angular resolution studies of all the $J = 2\rightarrow1$ lines.
ALMA high-resolution imaging  with 15 km long baselines, which was available for the first (2014) ALMA Long Baseline Campaign \citep{ALMALBC2015}, can deliver synthesized beams with FWHM values of tens of milliarcseconds. This will allow resolving the radio photosphere of the
nearby IRC+10216 and detailed imaging of the 0\as6 diameter circumstellar region
that contains vibrationally excited HCN, %(see Sect. \ref{hotcw}), 
while ALMA data previously published by \citet{Cerni2013} with $\approx 0\as6$ do not yet allow such imaging. We note that since the other stars of our sample are much more distant than IRC+10216, imaging with the longest ALMA baseline is necessary to provide interesting constraints on the innermost regions of their envelopes. Moreover, continuum emission from the radio photospheres, while easily detectable at 177 GHz, will require higher-frequency imaging to be resolved.

\begin{acknowledgements}
D. Keller was supported for this research by the International Max-Planck-Research School (IMPRS) for Astronomy and Astrophysics at the Universities of Bonn and Cologne and the Bonn-Cologne Graduate School (BCGS) for Physics and Astronomy. We thank Ankit Rohatgi for making his WebPlotDigitizer%r\footnote{http://arohatgi.info/WebPlotDigitizer/} 
tool available as open source software. We are grateful to Javier Alcolea for comments and to Yan Gong, Christian Henkel, and the referee for reading the manuscript and their valuable comments. The referee is thanked for a meticulous job.
\end{acknowledgements}

%\bibliographystyle{./aa}
%\bibliography{./KMM}
\bibliographystyle{aa}
\bibliography{KMM}

%\acknowledgments{This research has made use of the International Variable Star Index (VSX) database, operated at AAVSO, Cambridge, Massachusetts, USA.}
\begin{appendix}
\section{\label{appendix}Stellar phase determination}
In order to derive the IR phases corresponding to the date of APEX observations, we required the corresponding periods and dates of a IR maximum, $t_0$. While periods are relatively well known and published for most of our sources \citep{Whitelock2006}, $t_0$ is generally not listed in the literature. We used the IR photometry from \citet{Whitelock2006} to independently derive $t_0$ and the period for most of the APEX targets. Simple cosine curves were fitted using the least-squares method. Most derived periods were consistent with the published values; in cases when only few photometric points were available, the period was fixed at the value derived by \citet{Whitelock2006}.

For two sources, X Cnc and U Hya, we could not find literature or archival IR data and instead used the information provided by AAVSO\footnote{\url{https://www.aavso.org/vsx}} to derive their visual phases in the time of APEX observations. For most regular Mira variables, the IR light curve is delayed in  phase by about 0.1 with respect to the visual phase. The phase could not be determined for X Tra and X Vel owing to the lack of information on the period and/or times of maxima.

Because the variability analysis here is based on scarce photometric data from cycles long in the past and the light curves often display a non-periodic component, the derived phases should be considered as very rough estimates.
\end{appendix}
\end{document}